\documentclass[journal]{IEEEtran}
\usepackage{graphicx,amssymb,mathtools,cite}
\usepackage{mathrsfs,paralist,multirow,flushend}
\usepackage{epstopdf}
\usepackage{mathtools}
\begin{document}
\title{An E{ff}icient Approach for obtaining Feasible solutions from SOCP formulation of ACOPF}
\author{Anamika Tiwari, Abheejeet Mohapatra and Soumya Ranjan Sahoo
\thanks{Anamika Tiwari, Abheejeet Mohapatra and Soumya Ranjan Sahoo are with the Department of Electrical Engineering, Indian Institute of Technology Kanpur, 208016, India. (emails: anamtiw@iitk.ac.in, abheem@iitk.ac.in, srsahoo@iitk.ac.in)}}
\maketitle
\begin{abstract}
Exact Second Order Conic Programming (SOCP) formulation of AC Optimal Power Flow (ACOPF) consists of non-convex arctangent constraints. Generally, these constraints have been ignored or approximated (at the expense of increased computational time) so as to solve the relaxed and convex SOCP formulation of ACOPF. As a consequence, retrieving unique and feasible bus voltage phasors for ACOPF of meshed networks is not always possible. In this letter, this issue has been addressed. The arctangent constraints have been represented by alternate linear constraints in the relaxed SOCP formulation of ACOPF, by exploiting the properties of the meshed power network. Numerical tests show that the proposed formulation gives an unique and feasible ACOPF solution, which is practically realizable from system operation perspective and with global optimality feature, as compared to other works reported in the literature. Moreover, the proposed formulation is extremely efficient as only one execution of formulation provides a feasible solution.
\end{abstract}

\vspace{-1em}

\section{Introduction}
\IEEEPARstart{A}{COPF} is an operational problem in which the optimal power generation dispatch policy has to be determined within a stipulated time \cite{time7436515}. With the existing optimization solvers, convergence is not always guaranteed as ACOPF is a non-linear, non-convex, NP hard optimization problem. To achieve guaranteed convergence within the stipulated time, many ACOPF models have been proposed in the literature. Among these models, SOCP is quite popular due to its time efficiency to solve ACOPF \cite{reco} and the capability to provide global optimal solution for an exact convex formulation or a lower bound for a relaxed convex formulation \cite{bai2016robust}. SOCP formulation for power flow of meshed network has been first introduced in \cite{jabr2007conic}. The formulation in \cite{jabr2007conic} is non-convex due to the arctangent constraints on bus voltage phase angles. Generally, these constraints have been ignored in order to have a convex ACOPF formulation which can be used in multi-stage robust optimization \cite{bai2016robust}. In the absence of arctangent constraints, retrieving an unique set of bus voltage phasors for ACOPF of meshed network is not possible as the sum of difference of bus voltage phase angles in a mesh/ cycle is not equal to zero \cite{kocuk2016strong}. This renders the solution of robust optimization to be practically unrealizable \cite{reco}.

Alternatively, the arctangent constraints in ACOPF have also been approximated through appropriate constraints in \cite{reco,kocuk2016strong}. In \cite{reco}, feasible ACOPF solution from convex SOCP formulations, has been obtained. However, the approach in \cite{reco} is iterative and computationally intensive as each iteration requires solution of a distinct and convex SOCP formulation, which may not be desirable from the perspective of time involved in system operation \cite{time7436515}. Also, the convergence of approach in \cite{reco} highly depends on a penalty parameter and a good choice of initial solution. Further, the problem size in \cite{reco} increases significantly with the system size. In \cite{kocuk2016strong}, bilinear cyclic/ valid linear constraints on phase angles have been used to tighten the SOCP and Semi-Definite Programming (SSDP) formulations of ACOPF, which again requires solutions of distinct and multiple SOCP/ SDP formulations. Thus, approaches in \cite{kocuk2016strong} are also computationally intensive.Hence, a computationally efficient approach to generate unique and practically feasible ACOPF solutions for meshed networks using SOCP is missing in the literature. 

In this letter, a novel approach to efficiently find unique and globally optimal ACOPF solutions for meshed network from convex SOCP formulation is proposed. The obtained bus voltage phasors are practically feasible as well as physically implementable. For this, the arctangent constraints are represented through alternate linear constraints in the convex SOCP formulation by using the properties of the power network. Numerical tests prove the efficacy of the proposed formulation with the resultant optimality gap of the obtained solutions being close to zero for most of the test cases, as compared to other works reported in the literature.\vspace{-1em}

\section{SOCP formulation of ACOPF \cite{reco,jabr2007conic}}
Let the meshed power network be represented by $(\mathcal{N},\mathcal{B})$, where $\mathcal{N}$ is the set of buses and $\mathcal{B}\subseteq\mathcal{N}\times\mathcal{N}$ represents the set of branches. $\mathcal{G}\subseteq\mathcal{N}$ and $\mathcal{L}\subseteq\mathcal{N}$ denote the set of generator and load buses, respectively. ACOPF can be stated as
\begin{align}\label{1}
\min&\sum_{i\in\mathcal{G}}\left[a_iP_{gi}^2+b_iP_{gi}+c_i\right]\\\label{2}
\text{subject to }&P_{gi}-P_{di}=P_i;\forall i\in\mathcal{G}\\\label{3}
&-P_{di}=P_i;\forall i\in\mathcal{L}\\\label{221}
&Q_{gi}-Q_{di}=Q_i;\forall i\in\mathcal{G}\\\label{31}
&-Q_{di}=Q_i;\forall i\in\mathcal{L}\\\label{4}
&-\overline{P}_{ij}\leq P_{ij}\leq\overline{P}_{ij};\forall (i,j)\in\mathcal{B}\\\label{5}
&\underline{V}_i\leq V_i\leq\overline{V}_i;\forall i\in\mathcal{N}\\\label{6}
&\underline{P}_{gi}\leq P_{gi}\leq\overline{P}_{gi};\forall i\in\mathcal{G}\\\label{7}
&\underline{Q}_{gi}\leq Q_{gi}\leq\overline{Q}_{gi};\forall i\in\mathcal{G}\\\label{8}
&-\pi/2\leq \theta_i\leq\pi/2;\forall i\in\mathcal{N},\theta_{ref}=0
\end{align}
where, $a_i$, $b_i$ and $c_i$ are cost coefficients of the $i^{th}$ generator. $P_{gi}$ and $Q_{gi}$ are the associated real and reactive power generations, respectively. $P_{di}$ and $Q_{di}$ are the real and reactive power demand at bus $i$, respectively. At each bus $i \in\mathcal{N}$, the real and reactive power injections $P_i$ and $Q_i$, respectively are defined as
\begin{align}\label{81}
P_i=&\sum_{j\in M(i)}V_iV_j\left[G_{ij}\cos\theta_{ij}+B_{ij}\sin\theta_{ij}\right]+G_{ii}V_i^2 \\\label{82}
Q_i=&\sum_{j\in M(i)}V_iV_j\left[G_{ij}\sin\theta_{ij}-B_{ij}\cos\theta_{ij}\right]-B_{ii}V_i^2
\end{align}
where, $V_i\angle\theta_i$ is the voltage phasor of bus $i$.  $M(i)$ is the set of buses connected to bus $i$ and $\theta_{ij}=\theta_i-\theta_j$. $G_{ij}$ and $B_{ij}$ are the real and imaginary parts of $(i,j)^{th}$ element of admittance matrix, respectively. For each line $(i,j)\in\mathcal{B}$, $\overline{P}_{ij}$ is the associated maximum real power flow and $P_{ij}$ is the real power flow, which is
\begin{equation}\label{83}
P_{ij}=V_iV_j\left[G_{ij}\cos\theta_{ij}+B_{ij}\sin\theta_{ij}\right]-G_{ij}V_i^2
\end{equation}
$\underline{V}_i$/ $\overline{V}_i$, $\underline{P}_{gi}$/ $\overline{P}_{gi}$ and $\underline{Q}_{gi}$/ $\overline{Q}_{gi}$ are lower/ upper limits on bus voltage magnitude, real and reactive power generations, respectively. In \eqref{8}, $\theta_{ref}$ is phase angle of angle reference bus.

The original ACOPF is non-convex due to presence of $V_iV_j\cos\theta_{ij}$ and $V_iV_j\sin\theta_{ij}$ terms in \eqref{81} - \eqref{83}. Let, $c_{ii}:=V_i^2$, $c_{ij}:=V_iV_j\cos\theta_{ij}$ and $s_{ij}=V_iV_j\sin\theta_{ij}$. Then, \eqref{5}, \eqref{81} - \eqref{83} can be rewritten as
\begin{align}\label{511}
\underline{V}_i^2&\leq c_{ii}\leq\overline{V}_i^2;\forall i\in\mathcal{N}\\\label{811}
P_i&=\sum_{j\in N(i)}\left[G_{ij}c_{ij}+B_{ij}s_{ij}\right]+G_{ii}c_{ii};\forall i\in\mathcal{N}\\\label{821}
Q_i&=\sum_{j\in N(i)}\left[G_{ij}s_{ij}-B_{ij}c_{ij}\right]-B_{ii}c_{ii};\forall i\in\mathcal{N}\\\label{831}
P_{ij}&=G_{ij}(c_{ij}-c_{ii})+B_{ij}s_{ij};\forall (i,j)\in\mathcal{B}\\\label{16}
c_{ij}^2&+s_{ij}^2=c_{ii}c_{jj};\forall (i,j)\in\mathcal{B}
\\\label{17}
\tan&\theta_{ij}=s_{ij}/c_{ij};\forall (i,j)\in\mathcal{B}
\end{align}

The transformed ACOPF \eqref{1} - \eqref{4}, \eqref{6} - \eqref{8}, \eqref{511} - \eqref{17} is the exact SOCP formulation and is still non-convex due to \eqref{16} and \eqref{17}. In \cite{kocuk2016strong}, \eqref{16} is relaxed by its convex hull as
\begin{equation}\label{cone}
4c_{ij}^2+4s_{ij}^2+\left(c_{ii}-c_{jj}\right)^2\leq\left(c_{ii}+c_{jj}\right)^2;\forall (i,j)\in\mathcal{B}
\end{equation}
Arctangent constraints \eqref{17} are either ignored or approximated in \cite{reco,jabr2007conic,bai2016robust,kocuk2016strong} so as to have convex SOCP formulation of ACOPF. However by doing so, efficiently retrieving unique and practically feasible bus voltage phasors is difficult as the obtained solution, mostly, does not guarantee sum of bus voltage phase angle differences to be zero in every mesh of the network.\vspace{-1em}

\section{Proposed Formulation}
The key motivation behind this work is to efficiently retrieve bus voltage phasors that are feasible and physically realizable from SOCP formulation for ACOPF of meshed networks. For the obtained bus voltage phasors to be physically realizable, sum of bus voltage phase angle differences must be zero for every mesh in the network. In order to do so, the non-convex arctangent constraints \eqref{17} are represented through alternate linear constraints by using the properties of the power network. From \eqref{16} and \eqref{17}, it can be observed that
\begin{equation}\label{21}
\sin\theta_{ij}=\frac{s_{ij}}{\sqrt{c_{ii}c_{jj}}} = \frac{s_{ij}}{V_{i}V_{j}}
\end{equation}
Usually, under typical normal operating conditions of the power network, difference in bus voltage phase angles is close to zero and bus voltage magnitudes are close to $1$ pu \cite{reco}. Thus, \eqref{21} under such conditions can be rewritten as $\theta_i-\theta_j-s_{ij}\simeq 0$. This can be further relaxed as
\begin{equation}\label{23}
-\epsilon_\theta \leq \theta_i-\theta_j-s_{ij} \leq \epsilon_\theta 
\end{equation}
where, $\epsilon_\theta$ is a variable $\in[0,\overline{\epsilon_\theta}]$ with $\overline{\epsilon_\theta}=0.03491$ rad, which works well for typical operating conditions of the network.

With the arctangent constraints represented by linear constraints \eqref{23}, the new SOCP formulation for ACOPF of meshed networks can thus be stated as a convex formulation with the convex objective \eqref{1} and convex constraints \eqref{2}- \eqref{4}, \eqref{6} - \eqref{8}, \eqref{511} - \eqref{831}, \eqref{cone} and \eqref{23}.

 Unlike the approach in \cite{reco} and SSDP in \cite{kocuk2016strong}, the proposed formulation is easy to implement and gives solution in an extremely efficient manner, as only one execution of the formulation is required for obtaining a solution. Further, the obtained solution will be a globally optimal solution due to the convex formulation. Also, as bus voltage phase angles are optimization variables in the proposed formulation, sum of bus voltage phase angle differences is zero for every mesh in the network by use of \eqref{23}. This ensures that the obtained solution is practically feasible and can be physically realizable. Further, the number of cone constraints \eqref{cone} depend on cardinality of $\mathcal{B}$. Hence, the proposed new SOCP formulation is linearly scalable and can be efficiently solved using existing professional solvers.\vspace{-1em}

\section{Numerical Results}
The proposed formulation is tested on few NESTA test cases \cite{coffrin2014nesta} in MATLAB using SeDuMi on an Intel i7-2600 CPU with 3.40 GHz processor and 8 GB RAM. The obtained solution is compared with the same obtained from SOCP formulation without arctangent constraints \eqref{17} and SSDP formulation \cite{kocuk2016strong}. Percentage optimality gap of the solution is calculated as $\frac{Ob_{ac}-Ob_{.}}{Ob_{ac}}\times100$, where $Ob_{ac}$ is the objective of actual ACOPF and $Ob_{.}$ is the objective at the obtained solution. It is observed through various numeric results that solution with optimality gap close to zero, can be easily obtained when the proposed formulation is solved with objective \eqref{1} appended with $\beta\epsilon_\theta$, where $\beta=\sum_{i\in\mathcal{G}}\left(a_i\overline{P}_{gi}^2+b_i\overline{P}_{gi}+c_i\right)/\overline{\epsilon_\theta}$ is a weight for a given test case and hence, the same is used to obtain the following numeric results.

\begin{table*}
\begin{center}\vspace{-1em}
\caption{Percentage optimality gap and value of $\epsilon_\theta$ in the proposed new SOCP formulation for few NESTA test cases}
\label{tabel1}
\begin{tabular}{|c|c|c|c|c|c|c|c|c|}
\hline
& \multicolumn{4}{c|}{Typical Operating Conditions} & \multicolumn{4}{c|}{Congested Operating Conditions}\\\cline{2-9}
Test Case & SOCP & SSDP & new & $\epsilon_\theta$(rad) & SOCP & SSDP & new & $\epsilon_\theta$(rad)\\
& without \eqref{17} & \cite{kocuk2016strong} & SOCP & $(\times10^{-6})$ & without \eqref{17} & \cite{kocuk2016strong} & SOCP & $(\times10^{-6})$\\\hline
14ieee & 0.11 & 0.00 & 0.00 & 0.00216 & 1.35 & 0.00 & 0.96 & 0.00437\\\hline
30as & 0.06 & 0.00 & 0.00 & 0.0029 & 4.76 & 1.72 & 0.90 & 0.118\\\hline
30fsr & 0.39 & 0.03 & 0.00 & 0.0339 & 45.97 & 40.28 & 13.40 & 0.00514\\\hline
30ieee & 15.65 & 0.00 & 0.12 & 0.028 & 0.99 & 0.08 & 0.61 & 0.00437\\\hline
39epri & 0.05 & 0.01 & 0.01 & 0.596 & 2.99 & 0.00 & 0.92 & 1.46\\\hline
57ieee & 0.06 & 0.00 & 0.00 & 0.322 & 0.21 & 0.13 & 0.13 & 0.143\\\hline
118ieee & 2.10 & 0.25 & 0.03 & 0.370 & 44.19 & 39.09 & 10.72 & 0.412\\\hline
162ieee & 4.19 & 3.50 & 1.74 & 2.01 & 1.52 & 1.20 & 1.27 & 0.407\\\hline
300ieee & 1.19 & 0.30 & 0.22 & 5.87 & 0.85 & 0.15 & 0.05 & 7.58\\\hline
\end{tabular}
\end{center}\vspace{-2em}
\end{table*}

\begin{figure}[htbp]
\centering
\includegraphics[scale=0.34]{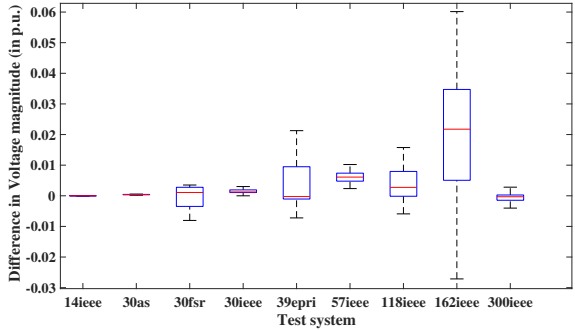}
\caption{Difference between voltage magnitude from proposed formulation and actual ACOPF solution under typical operating condition}
\label{Fig:delv}\vspace{-1em}
\end{figure}

Table \ref{tabel1} shows the percentage optimality gap of SOCP formulation without \eqref{17}, SSDP \cite{kocuk2016strong} and the proposed new SOCP formulation for the few NESTA test cases under typical and congested operating conditions as defined in \cite{coffrin2014nesta}. It is to be noted that the proposed formulation gives the solution in only one execution and is thus, extremely efficient as compared to approaches in \cite{reco,kocuk2016strong}. From Table \ref{tabel1}, it can be seen that the percentage optimality gap of solution from proposed formulation is close to zero for most of the test cases. Also, it is much less as compared to the same obtained from SOCP without \eqref{17} and SSDP under typical and congested operating conditions. The associated value of $\epsilon_\theta$ obtained in the proposed formulation is also given. It can be seen that $\epsilon_\theta$ is very small for all the test conditions and hence, value of $\beta\epsilon_\theta$ appended to \eqref{1} in the proposed formulation, is negligible as compared to the actual objective defined in \eqref{1}. Fig. \ref{Fig:delv} shows the difference between voltage magnitudes from proposed formulation and actual ACOPF solution under typical operating condition for all the test cases. Bus voltage magnitudes from proposed formulation are obtained as $V_i=\sqrt{c_{ii}};\forall i\in\mathcal{N}$. From Table \ref{tabel1} and Fig. \ref{Fig:delv}, it is clear that the proposed formulation gives solution which are close to the actual ACOPF solution and are indeed global due to the convex formulation.\vspace{-1em}

\begin{figure}[htbp]
\centering
\includegraphics[scale=0.3]{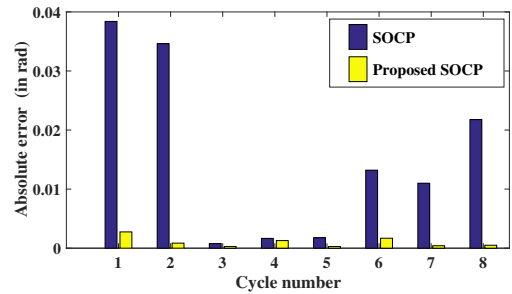}
\caption{Sum of phase angle differences for few cycles in 14ieee under congested operating condition}
\label{Fig:cycle}\vspace{-1.5em}
\end{figure}

For practical feasibility of the obtained solution, sum of all $\theta_{ij}$ over all meshes/ cycles of the test cases are evaluated using the solutions obtained from the proposed formulation. The same are also evaluated for SOCP without \eqref{17}, by explicit use of \eqref{17} after a solution from SOCP without \eqref{17}, is obtained. Fig. \ref{Fig:cycle} shows the sum of $\theta_{ij}$ for few cycles in 14ieee under congested operating condition from SOCP without \eqref{17} and proposed formulation. The associated cycles are 1) 1-2-5-1, 2) 2-3-4-2, 3) 2-4-5-2, 4) 6-12-13-6, 5) 6-11-10-9-14-13-6, 6) 5-6-11-10-9-4-5, 7) 1-5-6-12-13-14-9-7-4-3-2-1 8) 2-3-4-7-9-10-11-6-5-2. It can be seen that this sum is close to zero in the proposed formulation, whereas the same is not true for SOCP without \eqref{17}. This is also observed for all other test cases.

To check whether the solution obtained from proposed formulation is physically realizable or not, power flow is exclusively solved for all test conditions based on the converged solution of $P_{gi};\forall i\in{\mathcal{G}}-ref$, $V_i=\sqrt{c_{ii}};\forall i\in\mathcal{G}$ and $\theta_{ref}=0$ from proposed formulation for each test condition. It is observed that power flow converges and the associated inequality constraints evaluated from the power flow solution satisfy their respective limits for each case. This thus, ensures that the proposed formulation efficiently gives solutions which are practically feasible as well as physically realizable.\vspace{-1em}

\section{Conclusion}
A fully recoverable and convex SOCP formulation for the nonlinear, non-convex ACOPF problem is proposed here by representing the nonlinear arctangent/ cycle constraints through appropriate linear constraints. Solution obtained from the proposed formulation, when compared with the same from existing approaches in literature and actual ACOPF, prove that the proposed formulation is extremely efficient and is capable of giving globally optimal solutions which satisfy all power network constraints. This formulation is extremely easy to implement and hence, paves the way for using this in solving multi-stage robust optimization problems.
\bibliographystyle{IEEEtran}
\bibliography{IEEEabrv,ref}
\end{document}